\documentclass[epj,nopacs]{svjour}
\usepackage{axodraw}
\usepackage{epsfig}
\usepackage{amsfonts}
\usepackage{amsmath}
\usepackage{bbm}
\allowdisplaybreaks[1]

\input pix.sty

\renewcommand{\eq}{eq.~}
\renewcommand{\eqs}{eqs.~}
\renewcommand{\se}{sec.~}

\renewcommand{\fig}{fig.~}
\renewcommand{\figs}{figs.~}

\newcommand{\tinymsbar}{{\overline{\mbox{\tiny\rm{MS}}}}}
\newcommand{\Lambdamsbar}{{\Lambda_\tinymsbar}}

\newcommand{\Nf}{N_{\rm f}}
\newcommand{\Nv}{N_{\rm v}}
\newcommand{\Nc}{N_{\rm c}}

\newcommand{\Tc}{T_{\rm c}}

\newcommand{\rmO}{{\mathcal{O}}}
\newcommand{\bmu}{\bar\mu}

\def\lsi{\raise0.3ex\hbox{$<$\kern-0.75em\raise-1.1ex\hbox{$\sim$}}}
\def\gsi{\raise0.3ex\hbox{$>$\kern-0.75em\raise-1.1ex\hbox{$\sim$}}}

\newcommand{\gsim}{\mathop{\gsi}}

\newcommand{\nF}{n_\rmii{F}}

 \renewcommand{\nF}[1]{n_\rmii{F{#1}}}

\newcommand{\rmii}[1]{{\mbox{\tiny\rm{#1}}}}

\newcommand{\im}{\mathop{\mbox{Im}}}

\newcommand{\Tint}[1]{{\hbox{$\sum$}\!\!\!\!\!\!\!\int\,}_{\!\!\!\!\raise-0.9ex\hbox{$\scriptstyle{#1}$}}}
\newcommand{\Tinti}[1]{{{\Sigma}\!\!\!\!\raise0.3ex\hbox{$\int$}_\rmii{${#1}$}}}

\newcommand{\bi}{\begin{itemize}}
\newcommand{\ei}{\end{itemize}}

\renewcommand{\tt}{\tilde t_0}

\newcommand{\hide}[1]{ }

\newcommand{\f}{\mbox{\sl f\,}}
\newcommand{\ff}{\rmi{\sl f\,}}
\def\TAsc(#1,#2)(#3,#4,#5)%
{\SetWidth{2.0}\CArc(#1,#2)(#3,#4,#5)\SetWidth{1.0}}
\def\Lwidth{3}

\def\TAgl(#1,#2)(#3,#4,#5){\SetWidth{2.0}\PhotonArc(#1,#2)(#3,#4,#5){\Lwidth}%
{6.283 #3 mul 360 div #4 #5 sub #4 #5 sub mul sqrt mul Tdensity mul}%
\SetWidth{1.0}}
\def\TLgl(#1,#2)(#3,#4){\SetWidth{2.0}\Photon(#1,#2)(#3,#4){\Lwidth}
{#1 #3 sub #1 #3 sub mul #2 #4 sub #2 #4 sub mul add sqrt Tdensity mul}%
\SetWidth{1.0}}

\renewcommand{\picc}[1]{\;\parbox[c]{60pt}{\begin{picture}(60,30)(0,0)
\SetWidth{1.0}\SetScale{1.3} #1 \end{picture}}\; }
\def\Lwidth{1.3}

\makeatletter \@addtoreset{equation}{section} \makeatother

\begin{document}
\sloppy


\title{Towards flavour diffusion coefficient and
 electrical conductivity without ultraviolet contamination}
\titlerunning{Towards flavour diffusion coefficient and
 electrical conductivity}

\author{
Y.~Burnier\inst{1}
  \and 
M.~Laine\inst{2}
}

\institute{
Institute for Theoretical Physics, 
Albert Einstein Center, University of Bern, 
Sidlerstrasse 5, CH-3012 Bern, Switzerland
 \and
Faculty of Physics, University of Bielefeld, 
D-33501 Bielefeld, Germany}

\date{February 2012  \hfill BI-TP 2012/01}
 
\abstract{%
By subtracting from a recent lattice measurement of the thermal 
vector-current correlator the known 5-loop vacuum contribution, 
we demonstrate that the remainder is small and 
shows no visible short-distance 
divergence. It can therefore in principle be subjected to 
model-independent analytic continuation. Testing  
a particular implementation, we obtain estimates for the flavour-diffusion
coefficient ($2\pi T D \gsim 0.8$) and electrical conductivity  
which are significantly smaller than previous results. 
Although systematic errors remain beyond control
at present, some aspects of  
our approach could be of a wider applicability.
}

\maketitle


%
\section{Introduction}
\la{se:intro}

If a plasma is subjected to a perturbation whose wavelength is 
much longer than the typical 
mean-free path, then decohering scatterings take abundantly place 
within that perturbation, and its macroscopic physics should be
classical in nature. For instance, if we imagine
a hadronic ``jet'' with a width of a fermi or more, placed within  
a strongly interacting medium at a temperature $T \gg 200$~MeV, in 
which the characteristic distance scale is $1/(\pi T) \ll $~fm, 
then we expect to be able to describe the gross features of the 
jet's subsequent behaviour within classical hydrodynamics. The main role of 
quantum physics is that the classical description involves parameters, 
called transport coefficients, which need to be matched to the 
fundamental theory, in order to correctly capture the dynamics. 

As a particular case, let us consider 
a perturbation carrying a specific net quark flavour.  
In the absence of weak interactions, there is a conserved
current related to each flavour:  
$\partial_\mu J_{\ff}^\mu = 0$, $\f = 1, ..., \Nf$. The sum of all 
flavour currents (divided by the number of colours, $\Nc$) 
defines the baryon current, 
whereas a particular linear combination, weighted by the electric 
charges of all flavours, defines the electromagnetic current
(denoted by $J^\mu_\rmi{em}$). 

Now, to define the {\em flavour
diffusion coefficient}, denoted by $D^{ }_{\ff}$, requires  
a specification of the classical description onto which
to match. Let us assume that the 
perturbation is broad compared with $1/(\pi T)$, 
and express $J_{\ff}^\mu$ in 
a gradient expansion in $\sim \partial^\mu/\pi T$. Apart from $\partial^\mu$
the other Lorentz vector available is the four-velocity
defining the fluid rest frame ($u^\mu$; $u^\mu u_\mu = 1$). 
The free coefficients allowed by Lorentz 
symmetry are the transport coefficients. 
In particular, we can expand 
\be
 J_{\ff}^\mu =  u^\mu\,n^{ }_{\ff} + 
 D^{ }_{\ff}\, \partial^\mu_\perp n^{ }_{\ff} + \rmO(\partial^2) 
 \qquad \mbox{($\f$ fixed)}
 \;, \la{const_J}
\ee
where $n^{ }_{\ff}$ is the number density, and 
the transverse derivative has been defined as 
\be 
 \partial^\mu_\perp \equiv (\eta^{\mu\nu} - u^\mu u^\nu) \partial_\nu
 \;, \quad
 \eta^{\mu\nu} \equiv  \mathop{\mbox{diag}}\mbox{($+$$-$$-$$-$)}
 \;.  
\ee
The so-called Landau-Lifshitz convention 
is implied here, whereby
in the rest frame of the fluid the zero component of $J_{\ff}^\mu$ is nothing 
but the number density: $u_\mu J_{\ff}^\mu \equiv n^{ }_{\ff}$.

To match for $D^{ }_{\ff}$, it is helpful to specialize to simple
kinematics. In particular, 
going to the fluid rest frame and imposing current conservation 
on \eq\nr{const_J} yields
\be
 \partial^{ }_t\, n_{\ff} = D^{ }_{\ff} \nabla^2 n_{\ff} + \rmO(\nabla^3)
 \;. \la{Di_eq}
\ee 
In this frame it is 
a textbook exercise to derive an expression for $D^{ }_{\ff}$ in 
terms of 2-point correlators, which can then be matched onto 
quantum-mechanical expectation values; 
the result reads (cf.\ e.g.\ ref.~\cite{book}) 
\be
 D^{ }_{\ff} = 
 \fr1{3 \chi^{ }_{\ff}} 
 \lim_{\omega\to 0 } 
 \sum_{i=1}^{3}
 \frac{\rho_{ii}(\omega,\vec{0})}{\omega}
 \;, \la{Kubo_D}
\ee
where 
$
 \rho^{ }_{ii}(\omega,\vec{0}) \equiv \int_{t,\vec{x}} e^{i\omega t}
 \langle \fr12 [\hat{J}^i_{\ff}(t,\vec{x}) ,
 \hat{J}^i_{\ff}(0,\vec{0})  ]\rangle^{ }_T
$ 
is the spectral function related to  
the operators $\hat{J}^i_{\ff}$, and 
\ba
 \chi^{ }_{\ff}
 & \equiv & \int_0^\beta \! {\rm d}\tau \int_{\vec{x}}
 \Bigl\langle \hat J^0_{\ff}(\tau,\vec{x})
     \hat J^0_{\ff}(0,\vec{0}) \Bigr\rangle^{ }_T
 \nn 
 & = & \beta \int_{\vec{x}}
 \Bigl\langle \hat J^0_{\ff}(0,\vec{x})
   \hat J^0_{\ff}(0,\vec{0}) \Bigr\rangle^{ }_T
 \;, \quad
 \beta \equiv \frac{1}{T}
 \;, 
 \la{chi00}
\ea  
is the ``susceptibility'' related to the conserved charge. 
It is important to stress that even though this way of determining 
$D^{ }_{\ff}$ makes use of the fluid rest frame (reflected
by the vanishing spatial momentum in \eq\nr{Kubo_D}; in the following
this redundant argument is suppressed),  
the coefficient itself is defined also for relativistic flow; 
the corresponding covariant form of the diffusion equation 
is \eq\nr{const_J} together with $\partial_\mu J_{\ff}^\mu = 0$. 

In the case of the {\em electrical conductivity}, $\sigma$,
a possible definition is 
$
 \langle \hat{\vec{J}}_\rmi{em} \rangle = \sigma \vec{E}
$,
where $\vec{E}$ denotes an external electric field. 
Let us denote 
$
 D^{ }_{\ff} = D^{ }_{\ff,\rmi{s}} + D^{ }_{\ff,\rmi{ns}}
$, 
$
 \chi^{ }_{\ff} = \chi^{ }_{\ff,\rmi{s}} + \chi^{ }_{\ff,\rmi{ns}}
$, 
where 
$D^{ }_{\ff,\rmi{s}}$, 
$\chi^{ }_{\ff,\rmi{s}}$ are the contributions 
of the ``singlet'' or 
``disconnected'' quark contractions, 
and 
$D^{ }_{\ff,\rmi{ns}}$,
$\chi^{ }_{\ff,\rmi{ns}}$ those 
of the ``non-singlet'' or ``connected'' ones. 
One then obtains
\be
  \sigma =  e^2 \chi_{\ff}
  \Bigl[ 
  \Bigl( \sum_{\ff=1}^{\Nf} Q^{ }_{\ff} \Bigr)^2 
  D^{ }_{\ff,\rmi{s}}
  + 
  \Bigl( \sum_{\ff=1}^{\Nf} Q_{\ff}^2 \Bigr) 
  D^{ }_{\ff,\rmi{ns}}
 \Bigr]
 \;, \la{sigma}
\ee
where $Q^{ }_{\ff}$ denotes the electric charge of flavour $\f$ in units of
the elementary charge $e$. 

Although a lattice Monte Carlo determination of 
$D^{ }_{\ff}$ and $\sigma$ is numerically 
very demanding~\cite{harvey_rev}, 
a number of attempts have been 
launched in recent years~\cite{sg,aa,ding}. 
In particular, in ref.~\cite{ding} a continuum extrapolation of the 
relevant (connected) 
Euclidean correlator was carried out for the first time, with 
a philosophy that analytic continuation from the Euclidean correlator
to the Minkowskian spectral function
should only be attempted on the continuum-extrapolated
result. Subsequently various models were employed for the analytic
continuation, based on a few-parameter ansatz for $\rho^{ }_{ii}$. 
In the present work we make use of 
the continuum-extrapolated Euclidean
data of ref.~\cite{ding}, but analyze 
it in a different way, avoiding fits. 

The basic philosophy of our approach comes from ref.~\cite{cuniberti}, 
whose results were transcribed into a practical algorithm in 
ref.~\cite{analytic}. The main point is that in order to allow for 
an analytic continuation {\em in principle}, short-distance
divergences need to be subtracted from the Euclidean correlator, 
such that Fourier coefficients can be determined (in fact the function 
should even be continuous~\cite{cuniberti}). An important further 
insight comes from ref.~\cite{simon}, which showed that the
ultraviolet (UV) asymptotics
of the {\em thermal} contribution to the spectral function 
$\rho^{ }_{ii}$ is such that 
it does lead to a continuous Euclidean correlator. Therefore, only
the contribution of the 
{\em vacuum} $\rho^{ }_{ii}$ needs to be subtracted. The final ingredient
is that the vacuum $\rho^{ }_{ii}$ can be extracted from a recent 5-loop
computation of the vector current correlator~\cite{bck1}. Implementing
all these ingredients, we find surprisingly stable results which 
can be compared with other approaches, in order to obtain a rough impression
on the systematic uncertainties involved.

%
\section{Detailed setup}

We have in mind QCD with three massless valence flavours
($\Nv \equiv 3$). 
The gauge field configurations 
were generated within pure SU(3) gauge theory
in ref.~\cite{ding}, so the number of dynamical quarks
is zero ($\Nf = 0$). 
Moreover in ref.~\cite{ding} only the ``connected'' or ``non-singlet'' 
contractions were evaluated. For $\Nv = 3$
this implies that we are technically 
considering electric charge diffusion and susceptibility; following
ref.~\cite{ding} the corresponding coefficients are denoted 
by $D \equiv D^{ }_{\ff,\rmi{ns}}$ 
and $\chi^{ }_\rmi{q} \equiv \chi^{ }_{\ff,\rmi{ns}}$. 
Otherwise we keep the notation of 
\se\ref{se:intro} in the following, in particular continuing
to employ Minkowskian conventions for the Dirac matrices. 
(Note that the physics of {\em heavy flavour} diffusion, 
referring to quarks with a mass $M \gg \pi T$, 
is quite different from that of light quark diffusion~\cite{cst,eucl},
and the two cases should not be confused with each other, 
even though a single notation $D$ is often used 
for the diffusion coefficient; in particular, 
in the heavy quark case the extraction of $D$ might be somewhat 
more robust than here~\cite{kappaE,mumbai}.)

To specify the observables, it is helpful to start 
with the Lorentz covariant form of the 
vector current correlator, 
\ba
 G^{ }_{V}(\tau)
 & \equiv &
 - 
 \sum_{\mu = 0}^{3}
 \int_\vec{x} 
 \Bigl\langle (\bar\psi \gamma_\mu \psi) (\tau,\vec{x}) \;
 (\bar\psi \gamma^\mu \psi) (0,\vec{0})
 \Bigr\rangle^{ }_T
 \nn 
 & \equiv & 
 - G^{ }_{00}(\tau) + G^{ }_{ii}(\tau)
 \;, \la{GV_def}
\ea
where in the spatial part a sum over the indices is implied. 
Because of the projection to zero spatial momentum, 
$G^{ }_{00}(\tau)$ is actually $\tau$-independent, 
like in \eq\nr{chi00}; the value is denoted by 
$G^{ }_{00}(\tau) = \chi^{ }_\rmi{q} T$. It turns out that in the {\em free} 
limit, the spatial $G^{ }_{ii}(\tau)$ contains 
the same constant, which then cancels in the sum of \eq\nr{GV_def}~\cite{bie}. 
We denote this by $\chi_\rmi{q}^\rmi{free} = \Nc T^2/3$~\cite{av}. 
In the interacting theory, $G^{ }_{00}$ remains constant 
whereas $G^{ }_{ii}$ gets essentially modified.

The spectral functions corresponding 
to $G^{ }_V(\tau)$ and $G^{ }_{ii}(\tau)$ 
are denoted by $\rho^{ }_{V}(\omega)$ and $\rho^{ }_{ii}(\omega)$, 
respectively. If the spectral function is known, 
the Euclidean correlator can be obtained from 
\be
  G^{ }_{ii}(\tau) = 
 \int_0^\infty
 \frac{{\rm d}\omega}{\pi} \rho^{ }_{ii}(\omega)
 \frac{\cosh \left(\frac{\beta}{2} - \tau\right)\omega}
 {\sinh\frac{\beta \omega}{2}} 
 \;. \la{relation}
\ee
The basic issue is to what extent 
the inverse is true, i.e.\ information about the spectral
function, particularly concerning the transport coefficient 
$\lim_{\omega\to 0^+} \rho^{ }_{ii}(\omega)/\omega$ 
relevant for \eq\nr{Kubo_D}, can be 
extracted from a measured $G^{ }_{ii}$. 
An extensive
review on the problems encountered and the methods currently 
available can be found in ref.~\cite{harvey_rev}.

We note in passing that the relation in \eq\nr{GV_def} implies 
a corresponding relation of the spectral functions, 
$
 \rho^{ }_V(\omega) = \rho^{ }_{ii}(\omega)
  - \pi\chi^{ }_\rmi{q}\omega\delta(\omega)
$.
However the transport coefficients extracted from 
$\rho^{ }_V$ and $\rho^{ }_{ii}$ 
as $\lim_{\omega\to 0^+}\rho(\omega)/\omega$ are 
identical. In fact in the rigorous algorithm 
of ref.~\cite{cuniberti} the $\tau$-independent mode
gets explicitly projected out.  
 
Now, in ref.~\cite{simon}, 
the UV asymptotics of thermal spectral functions were
analyzed with Operator Product Expansion methods. In particular, it was
shown that thermal corrections to 
$\rho^{ }_{ii}(\omega)$ decrease at 
large frequencies as $\sim T^4/\omega^2$. When 
taken together with \eq\nr{relation} this statement, valid beyond 
perturbation theory, implies that the thermal part of $\rho^{ }_{ii}(\omega)$
yields a contribution to $G^{ }_{ii}(\tau)$ which is integrable even at
$\tau = 0$, i.e.\ remains finite at short distances. In contrast, the vacuum
spectral function yields a contribution diverging as $\sim 1/\tau^3$.
In order for the rigorous analytic continuation
of ref.~\cite{cuniberti} to be applicable, 
the divergent part needs to be 
subtracted~\cite{analytic}. 

In the literature, different strategies have been pursued
for the subtraction. In particular, in refs.~\cite{rhoE,Bulk_wdep}, 
analogous (but different) correlators were considered, and the idea
was to compute the whole $G^{ }_{ }(\tau)$, including thermal corrections, 
up to next-to-leading order  (NLO), i.e.\ 2-loop level, 
or $\rmO(\alpha_s)$. 
Although the same strategy
would also be possible for the vector correlator, 
by making use of classic results for the NLO thermal spectral
function~\cite{spectral1,spectral2,spectral3}, 
our goal here is to probe
a different strategy. Namely, the vacuum subtraction is handled with 
much {\em higher} precision than NLO, by making use of the fact 
that in vacuum, $\rho^{ }_{ii}(\omega)$ 
is known up to 5-loop level, or $\rmO(\alpha_s^4)$~\cite{bck1}.
In contrast, the thermal part is handled
with {\em lower} precision, only at leading order, leaving all other
thermal effects to be taken care of by the non-perturbative numerical
treatment of the remainder. 

Before proceeding it is important to underline once more the implications 
of the asymptotic behaviour $\sim T^4/\omega^2$~\cite{simon}.
In particular, the Lorentzian form 
\be
 \rho^\rmi{(L)}_{ii}(\omega) \equiv 3 D \chi^{ }_\rmi{q} \, 
 \frac{\omega \eta^2}{\omega^2 + \eta^2} 
 \la{Lorentz}
\ee
is sometimes used for modelling the transport peak. However, 
this shape can be correct only at {\em small frequencies}~\cite{amr}; 
at large frequencies $\omega \gg \eta$ it decays as 
$\sim 3 D \chi^{ }_\rmi{q}\eta^2/\omega$, which is slower than 
the mentioned $\sim T^4/\omega^2$ (and does not allow for 
a negative sign which is also a possibility~\cite{simon}). 
Most significantly, 
the Euclidean correlator obtained by inserting 
$\rho^\rmi{(L)}_{ii}$ into \eq\nr{relation} {\em diverges}
at $\tau \ll \beta$. So, if used as a part
of a fit ansatz for {\em all} $\omega$, 
this function may pick up an incorrect overlap 
on vacuum contributions, 
which could lead to an overestimate of~$D$.

%
\section{Vacuum spectral function}

We now turn to ref.~\cite{bck1} and specify the 5-loop vacuum 
spectral function. 
Following the conventions of ref.~\cite{bck2}, the 
coefficients of the $\beta$-function are defined according to 
\be
 \partial_t a_s = 
 - (\beta_0 a_s^2 + \beta_1 a_s^3 + \beta_2 a_s^4 + \ldots)
 \;, 
\ee
where 
\be
 a_s \equiv \frac{\alpha_s(\bmu)}{\pi}
 \;, \quad
 t \equiv \ln\biggl( \frac{\bmu^2}{\Lambda_\rmii{$\msbar$}^2} \biggr)
 \;, \la{Lam}
\ee
and, for $\Nc = 3$ (cf.\ ref.~\cite{beta}; we only need terms up to
the 3-loop level here),  
\ba
 \beta_0 & = & \frac{11}{4} - \frac{\Nf}{6}
  \;, \\ 
 \beta_1 & = & \frac{51}{8} - \frac{19\Nf}{24}
  \;, \\ 
 \beta_2 & = & \frac{2857}{128} - \frac{ 5033\Nf}{1152}
  + \frac{325\Nf^2}{3456}
  \;. 
\ea
The scale parameter $\Lambdamsbar$
in \eq\nr{Lam} represents an integration 
constant and, as usual, is chosen so that the asymptotic ($t \gg 1$) 
behaviour reads
\be
  a_s = \frac{1}{\beta_0 t} - \frac{\beta_1 \ln t}{\beta_0^3 t^2}
  + \frac{\beta_1^2 (\ln^2 t - \ln t - 1)  + \beta_2 \beta_0}
    {\beta_0^5 t^3} + \rmO\Bigl(\frac{1}{t^4}\Bigr)
 \;.
\ee

Then, the main result can be obtained from  
the coefficients $\Pi^0_i$, $i=0,...,4$, in \eqs(6)-(9) of 
ref.~\cite{bck2}. In vacuum, 
the quantity that we are interested in can be written as
\be
 \rho^{ }_{V}(\omega) = \frac{3\omega^2}{4\pi} \, R(\omega^2)
 \;, \la{rho_form}
\ee
where
\be
 R(\omega^2) = 4\pi 
 \im \Bigl\{\Pi^0_0 + \Pi^0_1 \, a_s + \Pi^0_2 \, a_s^2 + 
 \ldots \Bigr\}_{L \,\to\, \ell + i \pi}
 \;,
\ee
and 
\be
 \ell \equiv \ln\biggl(\frac{\bmu^2}{\omega^2}\biggr)
 \;. \la{ell_def}
\ee
Writing 
\ba
 R(\omega^2) & = &   
 r^{ }_{0,0} + 
 r^{ }_{1,0}\, a_s + 
 \bigl( r^{ }_{2,0} + r^{ }_{2,1}\, \ell \bigr)\, a_s^2 
   \nn & + & 
 \bigl( r^{ }_{3,0} + r^{ }_{3,1}\, \ell + r^{ }_{3,2}\, \ell^2 \bigr)\, a_s^3 
   \nn & + & 
 \bigl( r^{ }_{4,0} + r^{ }_{4,1}\, \ell + r^{ }_{4,2}\, \ell^2
  + r^{ }_{4,3}\, \ell^3\bigr)\, a_s^4
  + \rmO(a_s^5)
 \;, \nn \la{Rw}
\ea
the generalization of \eq(5) of ref.~\cite{bck1} to $\ell \neq 0$ reads 
\be
 \begin{array}{rcrcrcrcr}
 r^{ }_{0,0} & = & 1.0000000 & \;, & & &  \\[1mm]
 r^{ }_{1,0} & = & 1.0000000 & \;, & & &  \\[1mm]
 r^{ }_{2,0} & = &  1.9857074 & - & 0.1152954 \Nf &   \;, & &  \\[1mm]
 r^{ }_{2,1} & = &  2.7500000 & - & 0.1666667 \Nf &  \;, & &  \\[1mm]
 r^{ }_{3,0} & = &  -6.6369356 &  - & 1.2001341 \Nf &  - & 
 0.00517836 \Nf^2 & \;,  &  \\[1mm] 
 r^{ }_{3,1} & = &  17.296391  & - & 2.0876938 \Nf &  
  +&  0.03843180 \Nf^2 & \;,
 &  \\[1mm]
 r^{ }_{3,2} & = &  7.5625000 &  - & 0.9166667 \Nf & 
  + & 0.02777778 \Nf^2 & \;,
 &  \\[1mm] 
 r^{ }_{4,0} & = & -156.60811 &  + & 18.774765 \Nf &  - & 0.79743434 \Nf^2 & 
 \\[1mm] 
 & & & & &   + & 0.021516105 \Nf^3 & \;, \\[1mm] 
 r^{ }_{4,1} & = & -7.1166367 & -  & 15.565615 \Nf &  + & 0.83393599 \Nf^2 & 
 \\[1mm] 
 & & & & &   + & 0.002589178 \Nf^3 & \;, \\[1mm] 
 r^{ }_{4,2} & = &  88.878862 &  - & 16.175418 \Nf &  + & 0.81239907 \Nf^2 & 
 \\[1mm] 
 & & & & &   - & 0.009607950 \Nf^3 & \;, \\[1mm] 
 r^{ }_{4,3} & = & 20.796875 &  - & 3.7812500 \Nf &  + & 0.22916667 \Nf^2 & 
 \\[1mm] 
 & & & & &   - & 0.004629630 \Nf^3 & \;. \\[1mm]
 \end{array}   \la{coefflast}
\ee

As already mentioned, the lattice simulations of ref.~\cite{ding}
were for quenched QCD ($\Nf = 0$) and only evaluated the ``connected'' 
quark contraction. In the language of refs.~\cite{bck1,bck2} the latter
corresponds to ``non-singlet'' contributions, which are the only
ones included in the results above. Therefore, \eqs\nr{Rw}, \nr{coefflast}
with $\Nf = 0$ can directly be used for the analysis of the lattice 
data of ref.~\cite{ding}. (Ultimately singlet contributions will 
need to be included as well, and on the perturbative side 
progress in this direction is being made, 
cf.\ e.g.~ref.~\cite{singlet}.)

%
\section{Imaginary-time correlator}

\begin{figure}[t]


\centerline{%
 \epsfysize=7.5cm\epsfbox{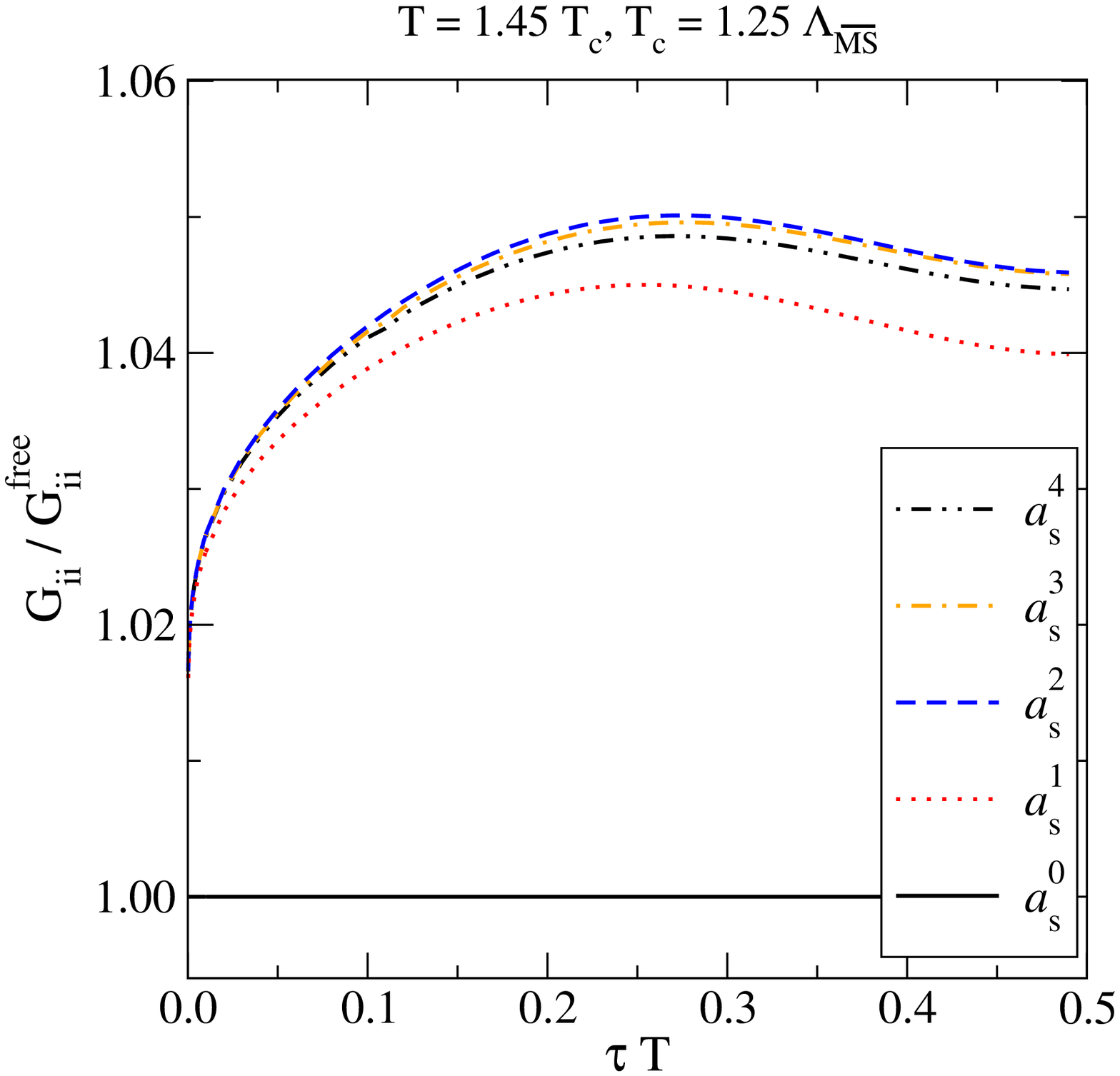}%
}

\caption[a]{\small
Five orders of $G^{ }_{ii}(\tau)$, from 
\eqs\nr{relation}, \nr{Rw}, \nr{rhoT_form}, normalized to 
the free result from \eq\nr{GVfree}. (The gauge coupling 
has always been evaluated with 3-loop running.)
}

\la{fig:GiiGiipert}
\end{figure}

Given the spectral function from \eq\nr{rho_form}, re-interpreted
as $\rho^{ }_{ii}$, we could 
insert it into \eq\nr{relation} and compute the corresponding
imaginary-time correlator. There is the problem, though, that 
technically arbitrarily small values of $\omega$ contribute 
to the integral, but for those the determination of $a_s$
becomes ill-defined. However, any modification of $\rho^{ }_{ii}$
in a finite range $0 \le \omega \le \omega_\rmi{max}$
yields a contribution to $G^{ }_{ii}(\tau)$
which does not diverge at small $\tau$ and can thus be taken 
care of together with the rest of the remainder. So, we cut 
the smallest frequencies off from $a_s$ by defining
\be
 \bmu_\rmi{ref} \equiv \mathop{\mbox{max}}(\pi T,\omega)
 \;, \la{muopt}
\ee
and by choosing $\bmu = \bmu_\rmi{ref}$ in the evaluation
of $G^{ }_{ii}(\tau)$. In addition we modify \eq\nr{rho_form}
by accounting for the leading-order thermal corrections: 
\be
 \rho_{ii}^{(T)}(\omega) \equiv \frac{3\omega^2}{4\pi} 
 \,
 \Bigl[ 1 - 2 \nF{}\Bigl( \frac{\omega}{2} \Bigr) \Bigr]
 \, R(\omega^2) + \pi \chi_\rmi{q}^\rmi{free} \omega\delta(\omega)
 \;, \la{rhoT_form}
\ee
where $\nF{}$ is the Fermi distribution. 
Following ref.~\cite{ding} the temperature is set to $T = 1.45\Tc$, 
and following e.g.\ ref.~\cite{bmw} we take $\Tc = 1.25\Lambdamsbar$
(uncertainties related to this choice are discussed below).
Like in ref.~\cite{ding} the results are 
normalized to the expression~\cite{ff}
\ba
 G_{ii}^\rmi{free}(\tau) & \equiv & 
 \int_{-\infty}^\infty
 \frac{{\rm d}\omega}{2\pi}
 \biggl\{ \frac{3\omega^2}{4\pi}
 \Bigl[ 1 - 2 \nF{}\Bigl( \frac{\omega}{2} \Bigr) \Bigr]
  + \pi T^2 \omega\delta(\omega)
 \biggr\}
 \nn & & \hspace*{2cm} \times \, 
 \frac{\cosh \left(\frac{\beta}{2} - \tau\right)\omega}
 {\sinh\frac{\beta \omega}{2}} 
 \nn & = & 
 6 T^3 \biggl[ 
  \pi (1-2\tau T) \frac{1 + \cos^2(2\pi\tau T)}{\sin^3(2\pi \tau T)}
 \nn & & \hspace*{2cm} + \,
 \frac{2 \cos(2\pi \tau T)}{\sin^2(2\pi \tau T)}
  + \fr16 \biggr]
 \;. \la{GVfree}
\ea
In \fig\ref{fig:GiiGiipert}, five subsequent orders of 
the perturbative result are shown;
in \fig\ref{fig:GiiGiilatt} the 5-loop perturbative result is 
compared with the continuum-extrapolated data from ref.~\cite{ding}. 

\begin{figure*}


\centerline{%
 \epsfysize=7.5cm\epsfbox{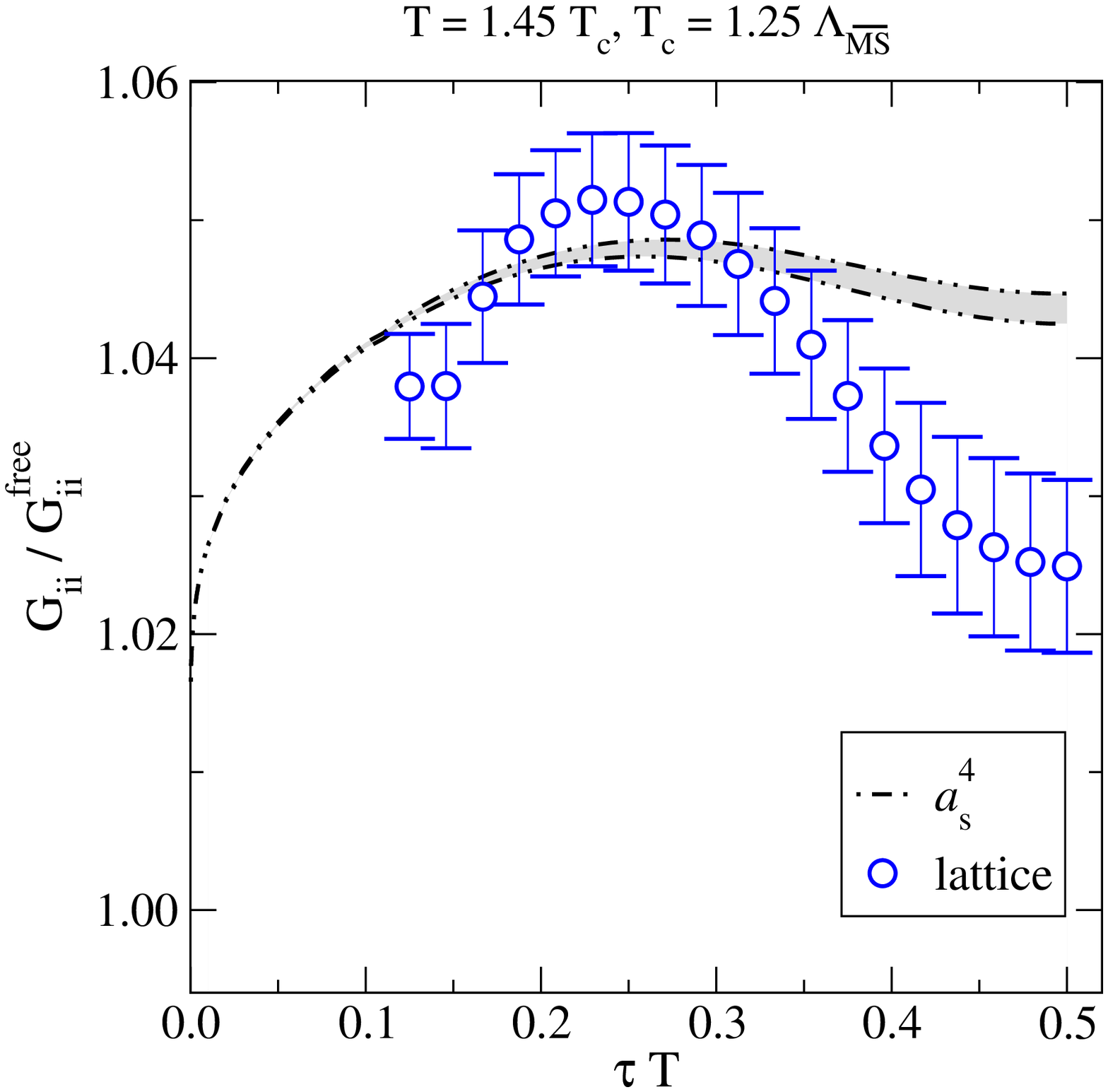}
~~~\epsfysize=7.5cm\epsfbox{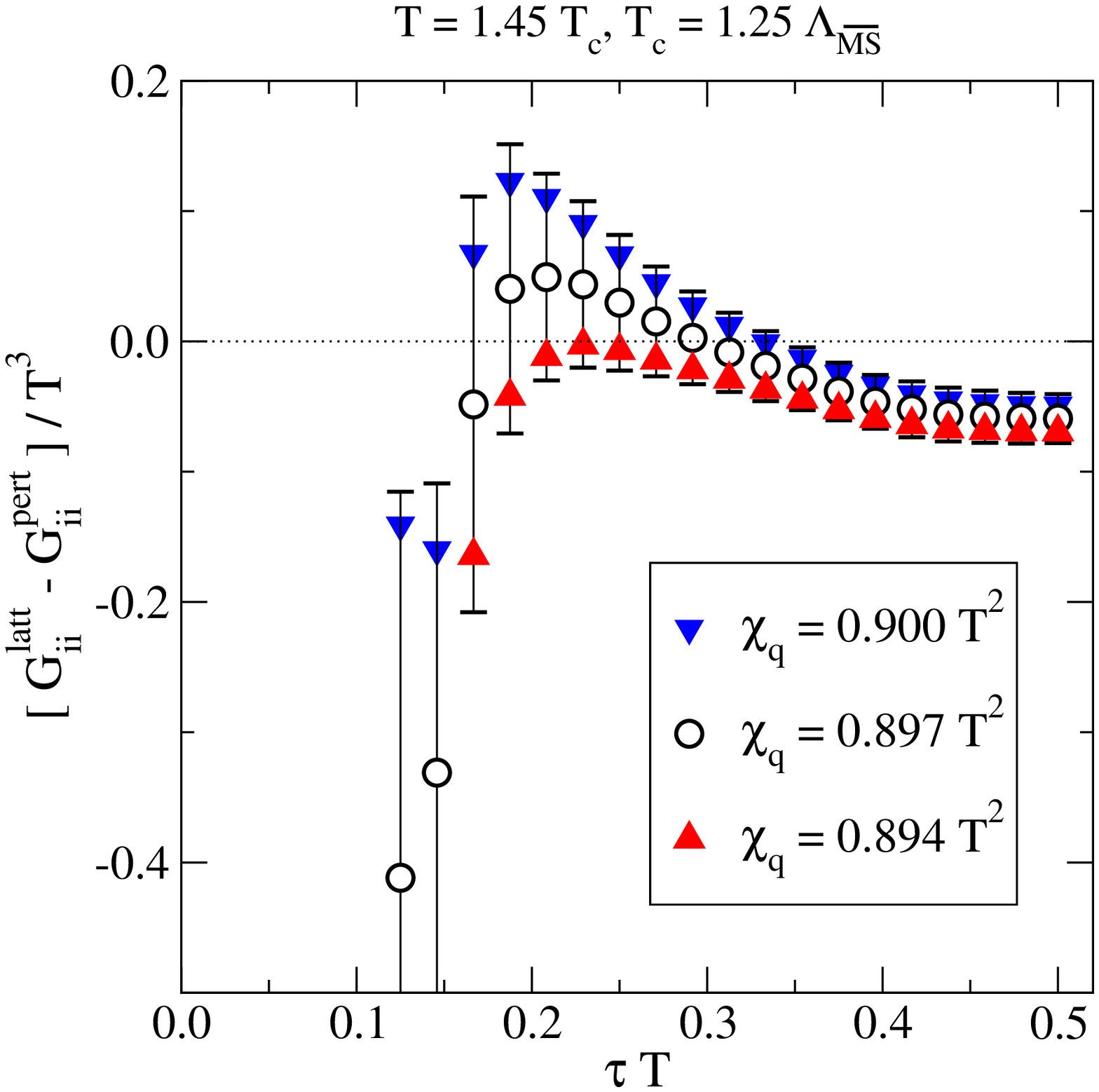}%
}

\caption[a]{\small
Left: Comparison of the $\rmO(a_s^4)$ result from 
\fig\ref{fig:GiiGiipert} with lattice data from ref.~\cite{ding}, 
multiplied by $\chi^{ }_\rmi{q}/T^2 = 0.897$. In the $\rmO(a_s^4)$ 
result, the renormalization scale has been varied within the range 
$\bmu = (1 ... 5) \times \bmu_\rmi{ref}$ (grey band).
Right: The difference of the lattice and perturbative results, 
with $\chi^{ }_\rmi{q}$ varied within its uncertainties.
Errors are shown for $\chi^{ }_\rmi{q}/T^2 = 0.897$ only. 
}

\la{fig:GiiGiilatt}
\end{figure*}

%
\section{Analytic continuation}

In order to estimate $D$ from \eq\nr{Kubo_D}, data for
the Euclidean $G^{ }_{ii}(\tau)$ should be fed into an analytic
continuation prescription, yielding the corresponding 
$\rho^{ }_{ii}(\omega)$. 
The function $G^\rmi{pert}_{ii}$ based
on \eq\nr{rhoT_form} yields a vanishing $D$, because
$\rho^{(T)}_{ii}$ has a vanishing slope around the origin. Therefore
we can equally well apply analytic continuation to the difference 
$G^\rmi{latt}_{ii} - G^\rmi{pert}_{ii}$. In fact, not only are 
we {\em allowed} to do this, but we probably {\em must} do this, 
given that to our knowledge mathematically justified 
analytic continuation to Minkowskian signature has been 
worked out only for a function which
has no singularity at small $\tau$~\cite{cuniberti}. 

\begin{figure}[t]


\centerline{%
 \epsfysize=7.5cm\epsfbox{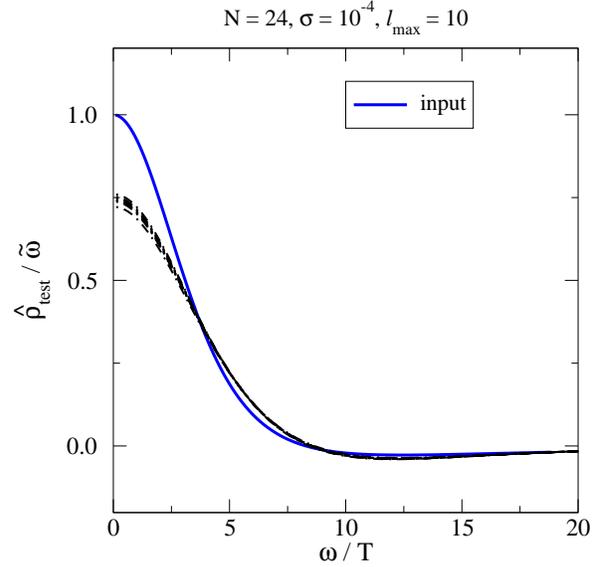}
}

\caption[a]{\small
A test of the method of ref.~\cite{cuniberti} with the 
example introduced in ref.~\cite{analytic}, employing 
an improved prescription. Here 
$\hat\rho$ refers to 
normalization by appropriate powers of $T$; and ``input'' 
denotes the correct result, known in this case.  
The result should be compared
with fig.~4(right) of ref.~\cite{analytic}, where an identical data
set lead to considerably more spread. 
}

\la{fig:rhotest}
\end{figure}

After the subtraction of the singular terms, we make use of the method 
of ref.~\cite{cuniberti}, as implemented in ref.~\cite{analytic}. 
In the meanwhile we have even found a significant improvement 
over the original implementation. 
The general method is based on determining
the Fourier coefficients from the Euclidean data $G(\tau)$, which 
can be done numerically, and using these in order to construct
an expansion of the real-time dependence of the correlator 
in terms of Laguerre polynomials, with coefficients denoted
by $a_\ell$. The physical real-time correlator must vanish at infinite
time separation~\cite{ay}, which corresponds to  
$
 \sum_{\ell = 0}^{\ell_\rmii{max}} a_\ell = 0
$. 
In practice, however, the correlator does not   
vanish exactly; choices of $\ell_\rmi{max}$
for which it vanishes approximately were dubbed 
``windows of opportunity'' in ref.~\cite{analytic}. Within
these windows, the small remaining asymptotic value was 
subtracted from the correlator by hand, 
in order to allow for a Fourier transform to Minkowskian frequency space.
We have now replaced the subtraction by another procedure; 
namely, the coefficient $a^{ }_{\ell_\rmii{max}}$ for which
the asymptotic value first crosses zero, is redefined to be 
\be
 a_{\ell_\rmii{max}}' \equiv - \sum_{\ell = 0}^{\ell_\rmii{max}-1} a_\ell
 \;. 
\ee
It turns out that this way the dependence on $\ell_\rmi{max}$
is much milder than with the original procedure; in fact
one does not even need to be close to a ``window'' but a plateau in 
$\sum_{\ell = 0}^{\ell_\rmii{max}} a_\ell^2$
would be sufficient as 
was envisaged in ref.~\cite{cuniberti}. 
The improvement is illustrated
in \fig\ref{fig:rhotest}, obtained for the same data set
(with the same simulated errors) as fig.~4 of ref.~\cite{analytic}; 
the results are substantially more stable, and equally close 
to the correct value (``input''), 
underestimating it by $\sim$ 25\%.  
(We have shown results for $N\equiv N_\tau =24$ to allow for a direct 
comparison with ref.~\cite{analytic}; the case $N=48$ 
corresponding to the resolution of ref.~\cite{ding} shifts 
the results by 2-3\% in the correct direction.)

\begin{figure*}


\centerline{%
 \epsfysize=7.5cm\epsfbox{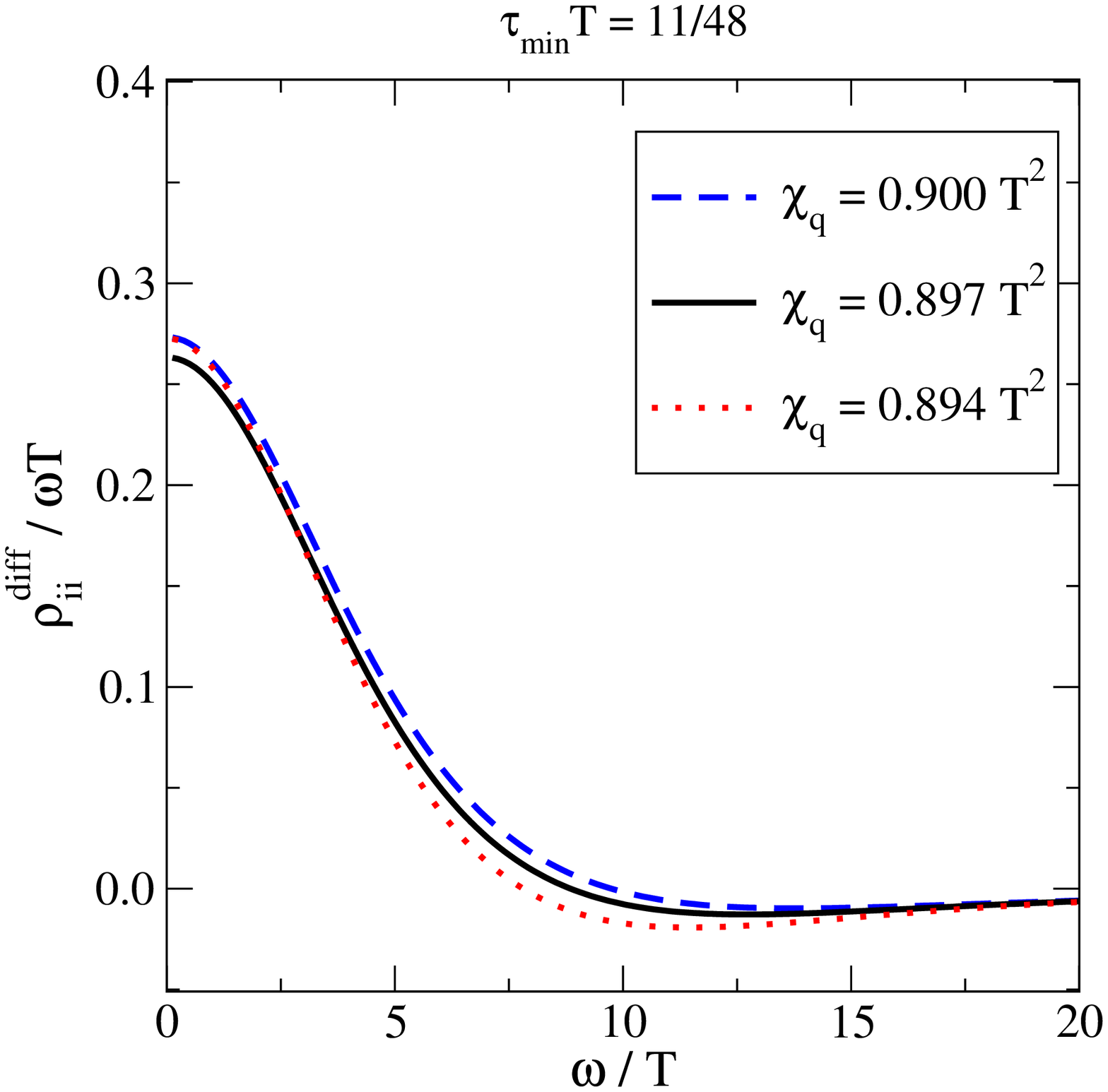}%
~~~ \epsfysize=7.5cm\epsfbox{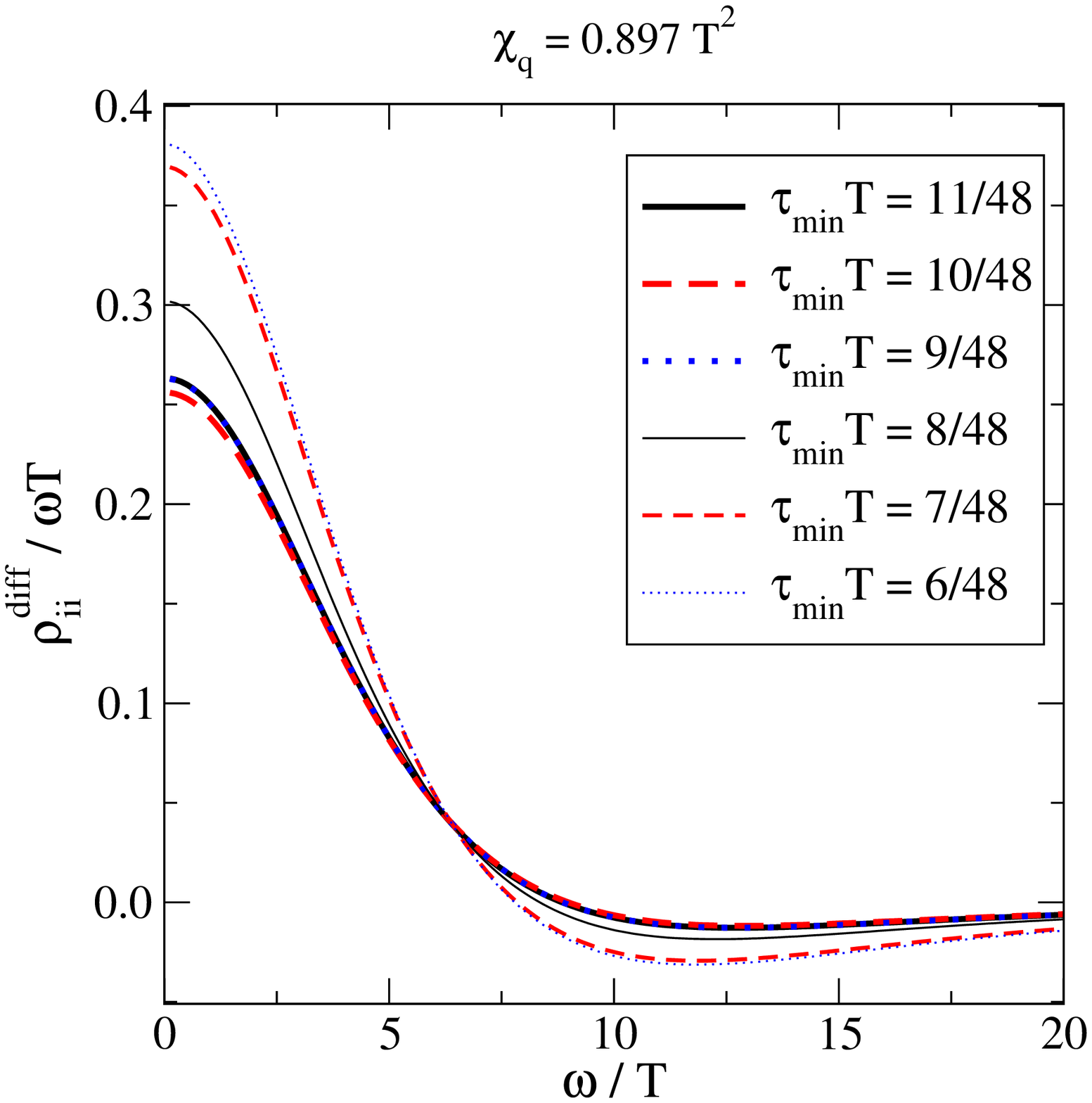}%
}

\caption[a]{\small
Left: Dependence of the spectral
function on $\chi^{ }_\rmi{q}/T^2$, for a fixed $\tau_\rmi{min} T = 11/48$.
Right: Dependence of the spectral 
function on $\tau_\rmi{min} T$, for a fixed $\chi^{ }_\rmi{q}/T^2 = 0.897$. 
}

\la{fig:rhodiff_errors}
\end{figure*}

Encouraged by this success, we have applied the algorithm 
to the difference 
$G^\rmi{diff}_{ii} \equiv G^\rmi{latt}_{ii} - G^\rmi{pert}_{ii}$ shown 
in \fig\ref{fig:GiiGiilatt}(right)
(the corresponding spectral function is 
denoted by $\rho^\rmi{diff}_{ii}$).\footnote{%
   The data for $G^\rmi{diff}_{ii}$ as well as for a corresponding
   $G^\rmi{diff}_{V}$ can be obtained from the authors on request. 
   The two differ by a constant mode; we stress that in the 
   algorithm of ref.~\cite{cuniberti} or in the fit described in 
   footnote~\ref{fit} an exactly constant mode has 
   no effect on the spectral function.  
 }  
At very short distances no lattice data exists, 
so the difference has been kept fixed at its value at
a chosen $\tau = \tau_\rmi{min}$
(note that there is no reason for the difference to vanish).
Various sources of systematic
errors have been probed. We have checked that 
variations of the renormalization scale, as indicated 
in~\fig\ref{fig:GiiGiilatt}(left), and variations of 
$\Tc/\Lambdamsbar$ within a range $1.25\pm 0.10$,
consistent with e.g.\ ref.~\cite{bmw}
(but also with earlier works, cf.\ sec.~4.2 of ref.~\cite{gE2}), 
have an effect 
smaller than variations of $\chi^{ }_\rmi{q}/T^2 = 0.897(3)$, 
to which the continuum-extrapolated results were normalized
in ref.~\cite{ding}. The errors from the latter variation
are shown in \fig\ref{fig:rhodiff_errors}(left). 
As can be anticipated from \fig\ref{fig:GiiGiipert},
using 3 or 4-loop vacuum results would only lead to
minor changes. Another source
of systematic errors is the continuum extrapolation of 
ref.~\cite{ding}; it might be prudent not to make use of  
results below  $\tau T \simeq 0.20$~\cite{ding}. 
In \fig\ref{fig:rhodiff_errors}(right)
we show the corresponding effects; in the following we 
restrict to $\tau_\rmi{min} T = 11/48$ which lies 
within a stable range.

\begin{figure}[t]


\centerline{%
 \epsfysize=7.5cm\epsfbox{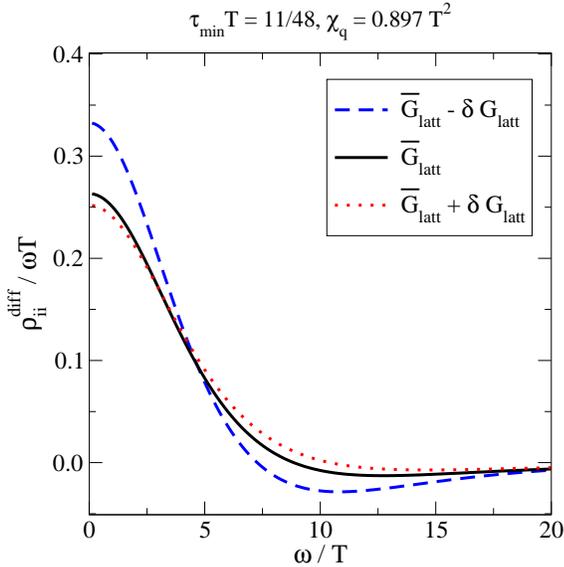}%
}

\caption[a]{\small
Results obtained by modifying lattice data 
by adding or subtracting its statistical errors.
}

\la{fig:rhostat}
\end{figure}

It remains to consider statistical errors. The errors as shown
in \fig\ref{fig:GiiGiilatt} are strongly correlated, but only 
available for each $\tau T$ separately. We have then 
shifted the whole function upwards or downwards by the errors. 
It is important to stress once again that 
the algorithm of ref.~\cite{cuniberti} exactly projects
out any constant contribution (i.e.\ Matsubara zero mode), 
so a uniform shift has no effect; the difference comes from 
the change in the shape. Results are shown in \fig\ref{fig:rhostat}; 
the variation from this rough implementation is smaller than
systematic errors related to the analytic continuation. 

%
\section{Conclusions}

Based on \figs\ref{fig:rhodiff_errors}, \ref{fig:rhostat}, 
and folding in an estimate of a downward systematic error
as suggested by \fig\ref{fig:rhotest}, we estimate
\be
 \lim_{\omega\to 0^+} \frac{\rho^{ }_{ii}(\omega)}{\omega}
 \; = \; 
 \lim_{\omega\to 0^+} \frac{\rho^\rmi{diff}_{ii}(\omega)}{\omega}
 \; \gsim\; 0.3\, T
 \;. \la{res_icpt}
\ee
Taking into account a factor 2 difference in the normalization
of $\rho^{ }_{ii}$, the corresponding estimate was cited as  
$(1 \ldots 3 ) T$ in ref.~\cite{ding}, i.e.\ 3 - 9 times larger
than our lower bound.
(In ref.~\cite{fk} only the 
lower limit was quoted, $D \chi^{ }_\rmi{q} \simeq 0.33 T$.
These estimates rely on \eq\nr{Lorentz} in combination with  
non-logarithmic modellings of the UV part of $\rho^{ }_{ii}$.\footnote{%
 If the remainder from \fig\ref{fig:GiiGiilatt}(right) 
 at $\tau T \ge 11/48$ is subjected to  
 a 3-parameter fit to \eq\nr{Lorentz} plus a $\tau$-independent
 constant, and errors are treated as uncorrelated, we obtain 
 $D \chi^{ }_\rmi{q} \simeq 0.30 T$, $\eta \simeq 1.1 T$, with 
 $\chi^2/\mbox{d.o.f.} \simeq 0.006$, numbers comparable 
 with ref.~\cite{fk}. The corresponding transport peak is markedly narrower
 than those in \figs\ref{fig:rhodiff_errors}, \ref{fig:rhostat}.
 Yet, as mentioned, the associated $G^\rmi{diff}_{ii}$  
 diverges at small $\tau$, whereas ours stays finite, 
 as required by ref.~\cite{simon}.
 If we cut off the large frequencies responsible for the divergence, 
 for instance by defining $\rho^\rmi{(L')}_{ii}(\omega) \equiv  
 \rho^\rmi{(L)}_{ii}(\omega)/\cosh(\frac{\omega}{2\pi T})$, 
 then the fit result moves in our direction:
 $D \chi^{ }_\rmi{q} \simeq 0.16 T$, $\eta \simeq 2.0 T$, with 
 $\chi^2/\mbox{d.o.f.} \simeq 0.005$.
 To resolve the correct physics    
 it hence appears important to reach a good resolution for
 the continuum extrapolation of $G_{ii}^\rmi{diff}$ also at small~$\tau$, 
 verifying that it saturates to a constant value there. \label{fit}
 }) 
From \eq\nr{Kubo_D}, 
inserting $\chi^{ }_\rmi{q} = 0.897 T^2$, \eq\nr{res_icpt} yields
\be
 2 \pi T D \; \gsim \; 0.8 
 \;.
\ee
The corresponding electrical conductivity from \eq\nr{sigma}
evaluates to $\sigma \,\gsim\, 0.07\, e^2 T$ for 
$\Nv = 3$, $\Nf = 0$.

These numbers are substantially smaller than those of the leading-order 
weak-coupling expansion~\cite{amy1,amy2}, but they are intriguingly 
close to the AdS/CFT suggestion $2\pi T D = 1$~\cite{kss}. 
However, the transport peak could be extremely narrow~\cite{mr} 
and therefore our results should be interpreted as 
lower bounds, as has already been indicated by the notation.
Continuum results are needed in the full $\tau$-range, 
and the resolution needs to be gradually increased, 
both in terms of statistical precision as well as in terms 
of $N_\tau$, in order to see whether the results 
stay put or show a slow evolution. (Unfortunately 
the functional dependences of the analytically continued
results on statistical variance and $N_\tau$ are 
not easily extracted~\cite{cuniberti}.) 

In any case, the principal feasibility of carrying 
out a model-independent short-distance vacuum subtraction
before attempting 
any analytic continuation or spectral modelling, thereby 
removing harmful UV contamination from the signal 
(cf.\ \fig\ref{fig:GiiGiilatt}), has hopefully become clear. 
This general philosophy can perhaps 
be applied to other correlators as well, 
for instance those related to 
components of the energy-momentum tensor, 
even if in that case temperature-dependent 
subtractions are needed in addition to the vacuum one~\cite{simon}. 

%
\section*{Acknowledgements}

 We thank H.-T.~Ding for helpful discussions and 
 providing us with lattice 
 data from ref.~\cite{ding}, 
 and O.~Kaczmarek for useful discussions.  
 Y.B.\ thanks the ITPP at Ecole Polytechnique 
 F\'ed\'erale de Lausanne, were
 part of this work was carried out, for hospitality.
 M.L.\ acknowledges partial support by the BMBF under project
 06BI9002.

%


\begin{thebibliography}{99}

\bibitem{book}
  J.I.~Kapusta and C.~Gale, 
  {\it Finite-Temperature Field Theory: Principles and Applications} 
  (Cambridge University Press, Cambridge, 2006).

\bibitem{harvey_rev}
  H.B.~Meyer,
  {\it Transport Properties of the Quark-Gluon Plasma: 
  A Lattice QCD Perspective,}
  Eur.\ Phys.\ J.\  A {47} (2011) 86
  [1104.3708].

\bibitem{sg}
  S.~Gupta,
  {\it The electrical conductivity and soft photon emissivity
  of the QCD plasma,}
  Phys.\ Lett.\  B {597} (2004) 57
  [hep-lat/0301006].

\bibitem{aa}
  G.~Aarts, C.~Allton, J.~Foley, S.~Hands and S.~Kim,
  {\it Spectral functions at small energies and the electrical conductivity in
  hot, quenched lattice QCD,}
  Phys.\ Rev.\ Lett.\  {99} (2007) 022002
  [hep-lat/0703008].

\bibitem{ding}
   H.-T.~Ding, A.~Francis, O.~Kaczmarek, 
   F.~Karsch, E.~Laermann and W.~Soeldner,
   {\em Thermal dilepton rate and electrical conductivity:
   An analysis of vector current correlation functions
  in quenched lattice QCD,}
  Phys.\ Rev.\  {D 83} (2011)  034504
  [1012.4963].

\bibitem{cuniberti}
  G.~Cuniberti, E.~De Micheli and G.A.~Viano,
  {\em Reconstructing the thermal Green functions at real times from those at
  imaginary times,}
  Commun.\ Math.\ Phys.\  {216} (2001) 59
  [cond-mat/0109175].

\bibitem{analytic}
  Y.~Burnier, M.~Laine and L.~Mether,
  {\it A test on analytic continuation of thermal imaginary-time data,}
  Eur.\ Phys.\ J.\  C {71} (2011) 1619
  [1101.5534].

\bibitem{simon}
  S.~Caron-Huot,
  {\it Asymptotics of thermal spectral functions,}
  Phys.\ Rev.\  D {79} (2009) 125009
  [0903.3958].

\bibitem{bck1}
  P.A.~Baikov, K.G.~Chetyrkin and J.H.~K\"uhn,
  {\em Hadronic $Z$- and $\tau$-Decays in Order $\alpha_s^4$},
  Phys.\ Rev.\ Lett.\  {101} (2008)  012002
  [0801.1821].

\bibitem{cst}
  J.~Casalderrey-Solana and D.~Teaney,
 {\em Heavy quark diffusion in strongly coupled $\mathcal{N}=4$ Yang-Mills,}
  Phys.\ Rev.\  D {74} (2006) 085012
  [hep-ph/0605199].

\bibitem{eucl}
  S.~Caron-Huot, M.~Laine and G.D.~Moore,
  {\em A Way to estimate the heavy quark thermalization rate from the lattice,}
  JHEP {04} (2009) 053
  [0901.1195].

\bibitem{kappaE}
  A.~Francis, O.~Kaczmarek, M.~Laine and J.~Langelage,
  {\em Towards a non-perturbative measurement of
   the heavy quark momentum diffusion coefficient,}
   PoS LATTICE{2011} (2011) 202
  [1109.3941].

\bibitem{mumbai} 
  D.~Banerjee, S.~Datta, R.~Gavai and P.~Majumdar,
  {\em Heavy Quark Momentum Diffusion Coefficient from Lattice QCD,}
  Phys.\ Rev.\ D {85} (2012) 014510
  [1109.5738].

\bibitem{bie}
  F.~Karsch, E.~Laermann, P.~Petreczky and S.~Stickan,
  {\em Infinite temperature limit of meson spectral functions
  calculated on the lattice,}
  Phys.\ Rev.\ D {68} (2003) 014504
  [hep-lat/0303017].

\bibitem{av}
  A.~Vuorinen,
  {\em Quark number susceptibilities of hot QCD up to $g^6 \ln g$,}
  Phys.\ Rev.\ D {67} (2003) 074032
  [hep-ph/0212283].

\bibitem{rhoE}
  Y.~Burnier, M.~Laine, J.~Langelage and L.~Mether,
  {\it Colour-electric spectral function at next-to-leading order,}
  JHEP {08} (2010) 094
  [1006.0867].

\bibitem{Bulk_wdep}
  M.~Laine, A.~Vuorinen and Y.~Zhu,
  {\em Next-to-leading order thermal spectral functions
  in the perturbative domain,}
  JHEP {09} (2011)  084
  [1108.1259].

\bibitem{spectral1}
  R.~Baier, B.~Pire and D.~Schiff,
  {\it Dilepton production at finite temperature: 
  Perturbative treatment at order $\alpha_s$,}
  Phys.\ Rev.\  D {38} (1988) 2814.

\bibitem{spectral2}
  Y.~Gabellini, T.~Grandou and D.~Poizat,
  {\it Electron-positron annihilation in thermal QCD,}
  Annals Phys.\  {202} (1990) 436.

\bibitem{spectral3}
  T.~Altherr and P.~Aurenche,
  {\it Finite temperature QCD corrections to lepton-pair 
  formation in a quark-gluon plasma,}
  Z.\ Phys.\  C {45} (1989) 99.

\bibitem{amr}
  G.~Aarts and J.M.~Mart{\'i}nez Resco,
  {\em Transport coefficients, spectral functions and the lattice,}
  JHEP {04} (2002) 053
  [hep-ph/0203177].

\bibitem{bck2}
  P.A.~Baikov, K.G.~Chetyrkin and J.H.~K\"uhn,
  {\em $R(s)$ and hadronic $\tau$-Decays 
  in Order $\alpha^4_s$: technical aspects,}
  Nucl.\ Phys.\ Proc.\ Suppl.\  {189} (2009) 49
  [0906.2987].

\bibitem{beta} 
  T.~van Ritbergen, J.A.M.~Vermaseren and S.A.~Larin,
  {\em The four-loop $\beta$-function in quantum chromodynamics,}
  Phys.\ Lett.\  {B 400} (1997)  379
  [hep-ph/9701390].


\bibitem{singlet}
  P.A.~Baikov, K.G.~Chetyrkin and J.H.~K\"uhn,
  {\it Adler Function, DIS sum rules and Crewther Relations,}
  Nucl.\ Phys.\ Proc.\ Suppl.\ \ {205-206} (2010) 237
  [1007.0478].

\bibitem{bmw}
  S.~Borsanyi, G.~Endrodi, Z.~Fodor, S.D.~Katz and K.K.~Szabo,
  {\em Lattice SU(3) thermodynamics and the onset of perturbative behaviour,}
  1104.0013.

\bibitem{ff}
  W.~Florkowski and B.L.~Friman,
  {\em Spatial dependence of the finite temperature
  meson correlation function,}
  Z.\ Phys.\  {A 347} (1994)  271.

\bibitem{ay}
  P.B.~Arnold and L.G.~Yaffe,
  {\em Effective theories for real-time correlations in hot plasmas,}
  Phys.\ Rev.\  D {57} (1998) 1178
  [hep-ph/9709449].

\bibitem{gE2}
  M.~Laine and Y.~Schr\"oder,
  {\it Two-loop QCD gauge coupling at high temperatures,}
  JHEP {03} (2005) 067
  [hep-ph/0503061].

\bibitem{fk}
  A.~Francis and O.~Kaczmarek,
  {\em On the temperature dependence of the electrical conductivity
  in hot quenched lattice QCD,}
  1112.4802.

\bibitem{amy1}
  P.B.~Arnold, G.D.~Moore and L.G.~Yaffe,
  {\em Transport coefficients in high temperature gauge theories.\ 
  1.\ Leading log results,}
  JHEP {11} (2000) 001
  [hep-ph/0010177].

\bibitem{amy2}
  P.B.~Arnold, G.D.~Moore and L.G.~Yaffe,
  {\em Transport coefficients in high temperature gauge theories.\ 
  2.\ Beyond leading log,}
  JHEP {05} (2003) 051
  [hep-ph/0302165].

\bibitem{kss}
  P.~Kovtun, D.T.~Son and A.O.~Starinets,
  {\it Holography and hydrodynamics: Diffusion on stretched horizons,}
  JHEP\ {10} (2003) 064
  [hep-th/0309213].

\bibitem{mr}
  G.D.~Moore and J.-M.~Robert,
  {\em Dileptons, spectral weights, and conductivity 
  in the quark-gluon plasma,}
  hep-ph/0607172.

\end{thebibliography}
\end{document}